An Accelerometer Based Instrumentation of the Golf Club:

Comparative Analysis of Golf Swings

Robert D. Grober
Department of Applied Physics
Yale University
New Haven, CT  06520

December 30, 2009

The motion of the golf club is measured using two accelerometers mounted at different points along the shaft of the golf club, both sensitive to acceleration along the axis of the shaft.  The resulting signals are resolved into differential and common mode components.  The differential mode, a measure of the centripetal acceleration of the golf club, is a reasonable proxy for club speed and can be used to understand details of tempo, rhythm, and timing.  The common mode, related to the acceleration of the hands, allows insight into the torques that generate speed in the golf swing.  This measurement scheme is used in a comparative study of twenty-five golfers in which it is shown that club head speed is generated in the downswing as a two step process.  The first phase involves impulsive acceleration of the hands and club.  This is followed by a second phase where the club is accelerated while the hands decelerate.  This study serves to emphasize that the measurement scheme yields a robust data set which provides deep insight into the tempo, rhythm, timing and the torques that generate power in the golf swing.

Introduction

The use of electronics in the shaft or club head of a golf club has been the subject of considerable past work [1]. Modern implementations offer a large number of sensors and computational power concealed within the shaft. Over time, the tendency has been to make ever more sophisticated measurements in an effort to obtain increasingly detailed understanding of the golf swing. This paper describes a relatively simple measurement which yields a remarkably robust data set.

Measurement and Signal Analysis Summary

The measurement is described in detail in the companion paper, "*An Accelerometer Based Measurement of the Golf Swing – Measurement and Signal Analysis*" [2]. In summary, two accelerometers are mounted in the shaft of a golf club with the sensing direction oriented along the axis of the shaft. One accelerometer is located under the grip, preferably at a point between the two hands. The other is located further down the shaft. The output of the accelerometers is digitized and broadcast wirelessly to a computer, enabling data storage and signal analysis. The data rate for the entire wireless system is 4.4 ms/cycle, each cycle yielding data from both accelerometers.

The data is analyzed within the context of the double pendulum model of the golf swing, as discussed in Jorgensen's *The Physics of Golf* [3]. The model is shown schematically in Fig. 1. Our implementation assumes no translational motion of the center of the swing and assumes all motion is confined to a plane. The upper arm of the pendulum models the arms and body as a rigid rod of length $l_0$ oriented at an angle $\theta$ with respect to the *x*-axis of the inertial, Cartesian, coordinate system fixed in the plane of the swing. The x-axis is aligned along the direction of gravitational acceleration. The golf club is modeled as a rigid rod of length $l_c$ at an angle $\phi$ with respect to the x-axis. The orientation of the golf club is traditionally measured by the angle $\beta = \theta - \phi$, which roughly corresponds to the angle through which the wrists are cocked. Note here that $\beta$ is measured in the opposite orientation from $\theta$ and $\phi$ and is consistent with the definition of Jorgensen.

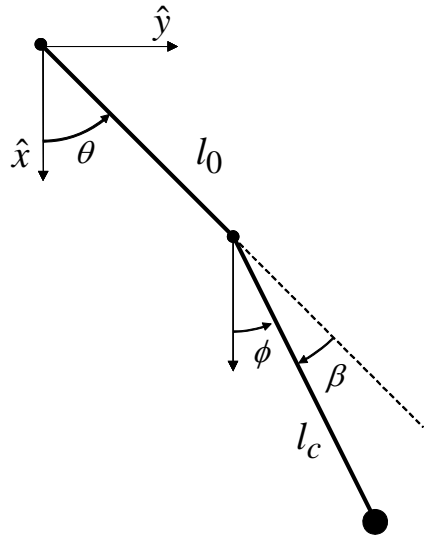

Fig 1. Geometry of the double pendulum. The angle $\theta$ defines the angle of the upper arm, $l_0$ with respect to the x-axis and defines the angle of the lower arm, $l_c$, with respect to the x-axis. The angle $\beta$ defines the angle of the lower arm with respect to the upper arm, and is interpreted as the wrist cocking angle.

The accelerometers are oriented along the axis of the golf club, their positions along the club measured from the hinged coupling of the two rods and given by the lengths $r_1$ and $r_2$. Their position in space is given as

$$\vec{r}_1 = (l_0 \cos\theta + r_1 \cos\phi)\hat{x} + (l_0 \sin\theta + r_1 \sin\phi)\hat{y} \tag{1a}$$

$$\vec{r}_2 = (l_0 \cos\theta + r_2 \cos\phi)\hat{x} + (l_0 \sin\theta + r_2 \sin\phi)\hat{y} \tag{1b}$$

Taking two derivatives, the generalized acceleration of the two points $\vec{r}_1$ and $\vec{r}_2$ is

$$\ddot{\vec{r}}_1 = -\left(l_0\dot{\theta}^2 \cos\theta + l_0\ddot{\theta}\sin\theta + r_1\dot{\phi}^2\cos\phi + r_1\ddot{\phi}\sin\phi\right)\hat{x}$$
$$-\left(l_0\dot{\theta}^2\sin\theta - l_0\ddot{\theta}\cos\theta + r_1\dot{\phi}^2\sin\phi - r_1\ddot{\phi}\cos\phi\right)\hat{y} \tag{2a}$$

$$\ddot{\vec{r}}_2 = -\left(l_0\dot{\theta}^2\cos\theta + l_0\ddot{\theta}\sin\theta + r_2\dot{\phi}^2\cos\phi + r_2\ddot{\phi}\sin\phi\right)\hat{x}$$
$$-\left(l_0\dot{\theta}^2\sin\theta - l_0\ddot{\theta}\cos\theta + r_2\dot{\phi}^2\sin\phi - r_2\ddot{\phi}\cos\phi\right)\hat{y} \tag{2b}$$

Rewriting in the $r$-$\phi$ coordinate system attached to the golf club,

$$\ddot{\vec{r}}_1 = -\left(r_1\dot{\phi}^2 + l_0\dot{\theta}^2\cos\beta + l_0\ddot{\theta}\sin\beta\right)\hat{r} + \left(r_1\ddot{\phi} - l_0\dot{\theta}^2\sin\beta + l_0\ddot{\theta}\cos\beta\right)\hat{\phi} \tag{3a}$$

$$\ddot{\vec{r}}_2 = -\left(r_2\dot{\phi}^2 + l_0\dot{\theta}^2\cos\beta + l_0\ddot{\theta}\sin\beta\right)\hat{r} + \left(r_2\ddot{\phi} - l_0\dot{\theta}^2\sin\beta + l_0\ddot{\theta}\cos\beta\right)\hat{\phi} \tag{3b}$$

Projecting the acceleration along the negative $\hat{r}$-axis yields a positive centripetal acceleration and the resulting signals on each accelerometer are

$$S_1 = -\hat{r} \cdot \ddot{\vec{r}}_1 = r_1\dot{\phi}^2 + l_0\dot{\theta}^2\cos\beta + l_0\ddot{\theta}\sin\beta + g^*\cos\phi \tag{4a}$$

$$S_2 = -\hat{r} \cdot \ddot{\vec{r}}_2 = r_2\dot{\phi}^2 + l_0\dot{\theta}^2\cos\beta + l_0\ddot{\theta}\sin\beta + g^*\cos\phi \tag{4b}$$

Gravitational acceleration has been added to these equations. Note that the magnitude of $g^*$ is the projection of the gravitational acceleration into the plane of motion, and is therefore less than the gravitational acceleration, $g = 9.8$ m/s$^2$.

$S_1$ and $S_2$ can be represented in terms of a common mode component, $F(t)$, and differential mode component, $G(t)$, defined as

$$F(t) = l_0\left(\dot{\theta}^2\cos\beta + \ddot{\theta}\sin\beta\right) + g^*\cos\phi \tag{5a}$$

$$G(t) = (r_1 - r_2)\dot{\phi}^2 \tag{5b}$$

Using these definitions, $S_1$ and $S_2$ can be written as,

$$S_1 = F(t) + \frac{r_1}{r_1 - r_2} G(t) \tag{6a}$$

$$S_2 = F(t) + \frac{r_2}{r_1 - r_2} G(t) \tag{6b}$$

The differential mode signal $G(t)$ is recovered from $S_1$ and $S_2$ by taking the difference of the two signals, $S_1 - S_2 = G(t) = (r_1 - r_2)\dot{\phi}^2$. For all the measurements presented in this paper, $r_1 - r_2 = 0.546$ m. As will be discussed below, this signal is a reasonable proxy for the speed of the club and can be used to provide insight into the tempo, rhythm, and timing of the golf swing. Its simplicity enables real-time biofeedback [4].

Determining $F(t)$ from $S_1$ and $S_2$ is more problematic. Ref [2] proposes a solution in which the function $S_i(t)$ is written as $S_i(t) = F(t) + a_i G(t)$ and the quantity $a_i$ is determined by minimizing $\int dt [S_i(t) - a_i G(t)]^2$. This algorithm yields the definition

$$a_i = \frac{\int dt\, S_i(t) G(t)}{\int dt\, G(t)^2} \tag{7}$$

and the resulting expression $F(t) = S_i(t) - a_i G(t)$. This method is used throughout this paper. As will be discussed below, $F(t)$ is related to the acceleration of the hands and provides insight into the torques that generate speed in the golf swing.

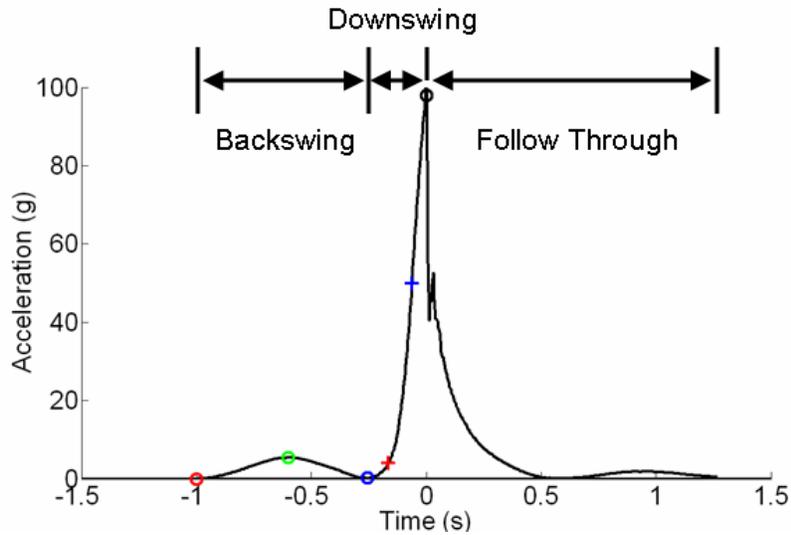

Fig 2(a). The differential mode signal $G(t)$, as described in the text, averaged over seven swings of a professional golfer. Impact is centered at $t = 0$. The y-axis is normalized to units of $g = 9.8$ m/s$^2$. The separation between sensors, $r_1$-$r_2 = 0.546$ m. The red circle is the start of the swing; green circle is the peak during the backswing; blue circle is the transition from backswing to downswing; and the black circle indicates impact. The red and blue crosses are defined in Fig. 2(b). The anomaly just to the right of impact is due to shock induced oscillations in the output of the sensor due to impact.

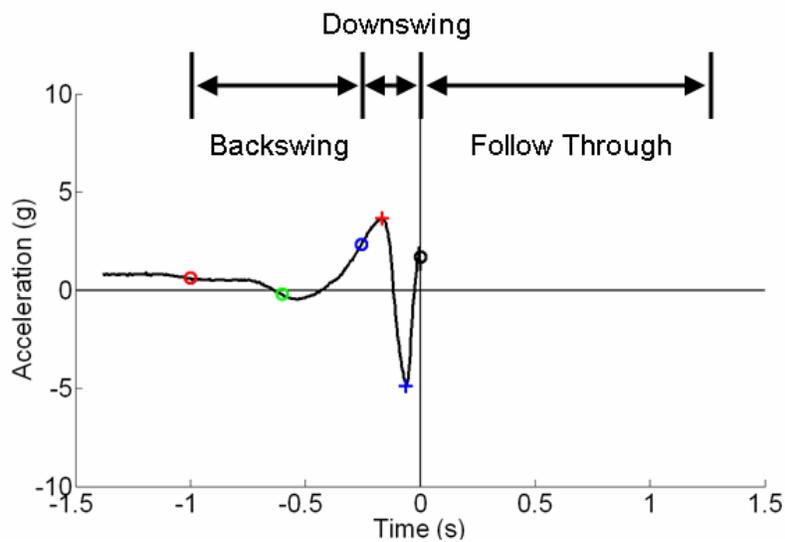

Fig 2(b): The common mode signal, $F(t)$, as described in the text, averaged over seven swings of a professional golfer. Impact is centered at $t = 0$. The y-axis is normalized to units of $g = 9.8$ m/s$^2$. The red cross is the peak of $F(t)$ and the blue cross is the minimum of $F(t)$. The open circles are defined in Fig. 2(a). We show the signal only until impact. The signal immediately after impact is dominated by shock induced oscillation of the sensor signals, which is not relevant to this paper.

Differential Mode Data Interpretation

Shown in Fig 2(a) is $G(t)$ averaged over seven swings of a professional golfer. The magnitude of the signal is normalized to the acceleration of gravity, $g$. Because the differential mode signal is a direct measure of the motion of the club, there are several points in the swing which are easily identified. They include 1) the beginning of the swing (red circle), 2) point of maximum speed in the backswing (green circle), 3) transition between backswing and downswing (blue circle), and 4) impact (black circle). Note that impact results in a large shock to the system, which drives oscillations in the sensor output. When averaged over several swings, this shock appears as the anomaly seen just to the right of the black circle in Fig. 2(a).

As an example of the usefulness of $G(t)$, one can determine the duration of the backswing and downswing for each individual swing, and then calculate the mean and standard deviation of the entire data set. For the data of Fig. 2(a), the average duration of the backswing is $731 \pm 21$ ms, and the average duration of the downswing is $258 \pm 8$ ms. The resulting ratio of backswing to downswing time is 2.8:1. This is very nearly the ratio of 3:1 first discussed by John Novosel in the book *Tour Tempo* [5], and subsequently by Grober, *et al*. [6], in which Novosel's observations were confirmed and a biomechanical explanation for the ratio was hypothesized.

Common Mode Data Interpretation

Shown in Fig. 2(b) is $F(t)$ calculated for the same data set shown in Fig. 2(a). The data is the average over seven swings of the player and is normalized to the acceleration of gravity, $g$. The four points measured in $G(t)$ (i.e. red, green, blue, and black circles) are indicated in Fig 2(b) so that the key points of the swing are easily identified. The data is only graphed thru impact, as shock induced oscillations dominate the region just after impact and yield no particularly useful information. An interesting feature in $F(t)$ is the structure observed during the downswing, characterized by the local maximum (red cross), which occurs very near the beginning of the downswing, and the local minimum (blue cross), which occurs when $G(t)$ is about half way to its maximum. The position of these features relative to $G(t)$ is indicated in Fig 2(a). The following paragraphs address the physical significance of these features.

As was discussed above, when interpreted in terms of the double pendulum $F(t)$ has the definition $F(t) = l_0\left(\dot{\theta}^2 \cos\beta + \ddot{\theta}\sin\beta\right) + g^* \cos\phi$. Because $F(t)$ has peaks that are several times larger than $g$, and since $g^*$ is less than $g$, the $g^*$ term can safely be ignored in the following analysis. Additionally, for most of the downswing the hands are cocked, $\beta$ is of order $\pi/2$, $\cos\beta \sim 0$ and $\sin\beta \sim 1$. Therefore, $F(t)$ can be approximated as $F(t) \sim l_0 \ddot{\theta}$, which is a measure of the acceleration of the point at which the two arms of the double pendulum are hinged. This is essentially the acceleration of the hands. Thus, the peak in $F(t)$ near the red cross corresponds to a region of acceleration of the hands, and the negative valued minimum near the blue cross corresponds to a deceleration of the hands.

To better understand why the hands accelerate and decelerate as observed in Fig. 2(b), one needs to understand the applied torque. To understand the torque one needs to consider the equations of motion of the double pendulum. The Lagrangian of the double pendulum is described in L.D. Landau and E.M. Lifschitz, *Mechanics* [7] and was first applied to the golf swing by Jorgensen [3]. Written in terms of the coordinate system of Fig. 1, the Lagrangian is

$$L = \frac{1}{2}\left(m_1 <l_0^2> + m_2 l_0^2\right)\dot{\theta}^2 + \frac{1}{2}m_2 <l_c^2> \dot{\phi}^2 + l_0 <l_c> \dot{\theta}\dot{\phi}\cos(\theta - \phi) + \tau_\theta \theta + \tau_\beta(\theta - \phi), \quad (8)$$

where $m_1$ is the mass of the upper arm, $l_0$ is the length of the upper arm, $<l_0^2>$ is the second moment of the upper arm, $m_2$ is the mass of the lower arm, $l_c$ is the length of the lower arm, $<l_c>$ is the first moment of the lower arm, and $<l_c^2>$ is the second moment of the lower arm. All moments are taken by averaging over the distribution of mass and are measured relative to their most proximal point. $\tau_\theta$ is an applied torque that works to increase the angle $\theta$ and $\tau_\beta$ is an applied torque that works to increase the angle $\beta = \theta - \phi$. Using the shorthand notation,

$$I_\theta = m_1 <l_0^2> + m_2 l_0^2, \quad I_\phi = m_2 <l_c^2> \quad \text{and} \quad S = m_2 <l_c> l_0, \quad (9)$$

the Lagrange equations of motion are

$$I_\theta \ddot{\theta} + S\ddot{\phi}\cos\beta + S\dot{\phi}^2 \sin\beta = \tau_\theta + \tau_\beta \quad (10a)$$

$$S\ddot{\theta}\cos\beta + I_\phi \ddot{\phi} - S\dot{\theta}^2 \sin\beta = -\tau_\beta \quad (10b)$$

Once again, we are concerned here with that region of the swing starting from the transition and going into the downswing, during which time the hands are cocked, $\beta$ is of order $\pi/2$, $\cos\beta \sim 0$ and $\sin\beta \sim 1$. Additionally, the torque $\tau_\beta$ is negative in the downswing because it acts so as to un-cock the wrists, reducing $\beta$. For convenience of notation, we define the release torque $\tau_R = -\tau_\beta$. The equations of motion are then simplified to

$$I_\theta \ddot{\theta} = \tau_\theta - \tau_R - S\dot{\phi}^2 \qquad (11a)$$

$$I_\phi \ddot{\phi} = \tau_R + S\dot{\theta}^2 \qquad (11b)$$

These two equations are particularly insightful and their physical relevance is the subject of the following paragraphs.

As is indicated in Eq. 11(a), the only torque that can drive $\ddot{\theta}$ positive is $\tau_\theta$, the torque generated by the body. The greater the applied torque, the greater is the acceleration of the hands, and the larger is this positive going region of the common mode signal. Because the angle $\beta$ does not change appreciably during this region, both the hands and the club must experience comparable acceleration. $\ddot{\phi}$ and $\ddot{\theta}$ are comparable at low speeds if $\tau_R \sim \dfrac{I_\phi}{I_\theta + I_\phi} \tau_\theta$. Values of $I_\phi$ are typically of order 0.2 kg m$^2$ while $I_\theta$ is typically of order 1.2 kg m$^2$ [8], which means $\tau_R << \tau_\theta$.

Thus, *the local maximum of the common mode signal at the beginning of the downswing is a measure of the acceleration of the hands and club in response to the torque generated primarily by the body, $\tau_\theta$.*

The common mode signal increases, goes through a maximum, and then quickly becomes negative. During this region where the common mode signal goes negative, the differential mode signal, which is proportional to $\dot{\phi}^2$, increases dramatically, which means $\ddot{\phi}$ is large and positive. As can be seen in Eq. 11(b), $\ddot{\phi}$ is driven both by the torque $\tau_R$, which is applied by the golfer in an effort to release the club, and by the centripetal term $S\dot{\theta}^2$, which is associated with hand speed. In contrast, and as is seen in Eq. 11(a), the torque $\tau_R$ and the centripetal term $S\dot{\phi}^2$ serve to decelerate the hands.

When $\tau_R$ and $S\dot{\phi}^2$ combine to be larger than $\tau_\theta$, $\ddot{\theta}$ goes negative. Thus, the negative going region of the common mode signal is a direct measure of the deceleration of the hands, which happens in response to the acceleration of the club during the release. *The greater the torque that releases the club and the faster the club moves, the more the hands decelerate and the larger the negative going dip in the common mode signal.*

Eventually the common mode signal goes through a local minimum. This generally occurs one-half to two-thirds of the way from the beginning of the downswing to impact. After this point, the common mode signal increases, eventually becoming positive very near to impact. The wrists fully un-cock during this later region of the swing, $\beta$ trends towards zero, $\cos\beta \sim 1$ and $\sin\beta \sim 0$. Thus, near to impact, the common mode signal $F(t) = l_0(\dot{\theta}^2 \cos\beta + \ddot{\theta} \sin\beta) + g^* \cos\phi$ is dominated by the centripetal acceleration $l_0 \dot{\theta}^2$ and by the gravitational acceleration $g^* \cos\phi$ both of which are positive definite.

In summary, the common mode signal typical of the golf swing of professional golfers has a distinctive structure during the downswing. It exhibits a positive valued maximum followed by a negative valued minimum, and finally becomes slightly positive just as the club gets to impact. The positive maximum is associated with the impulsive acceleration of the hands and club at the beginning of the downswing. The negative minimum is associated with the release of the club, which serves to decelerate the hands. *The greater the positive maximum, the more powerful is the acceleration at the start of the downswing. The deeper the negative minimum, the more powerful is the release of the club.*

Detailed Comparison of Five Golfers

Figs. 3 thru 7 are the differential and common mode signals for five very different golfers. The first is an elderly, avid, amateur golfer with a handicap of order twenty-five, the second is an avid, amateur golfer with a handicap of order ten, the third is a competitive collegiate golfer, the fourth is a PGA tour professional, and the fifth is a professional long drive competitor. In each case, the peak of $G(t)$ provides an indication of how fast the club is moving while $F(t)$ provides some indication for how club head speed is developed. Commentary for each data set is provided in the figure captions. These figures are meant to provide some insight as to how the data evolves from the case of an avid golfer who does not hit the ball very far to the case of a golfer who hits the ball just about as far as anybody.

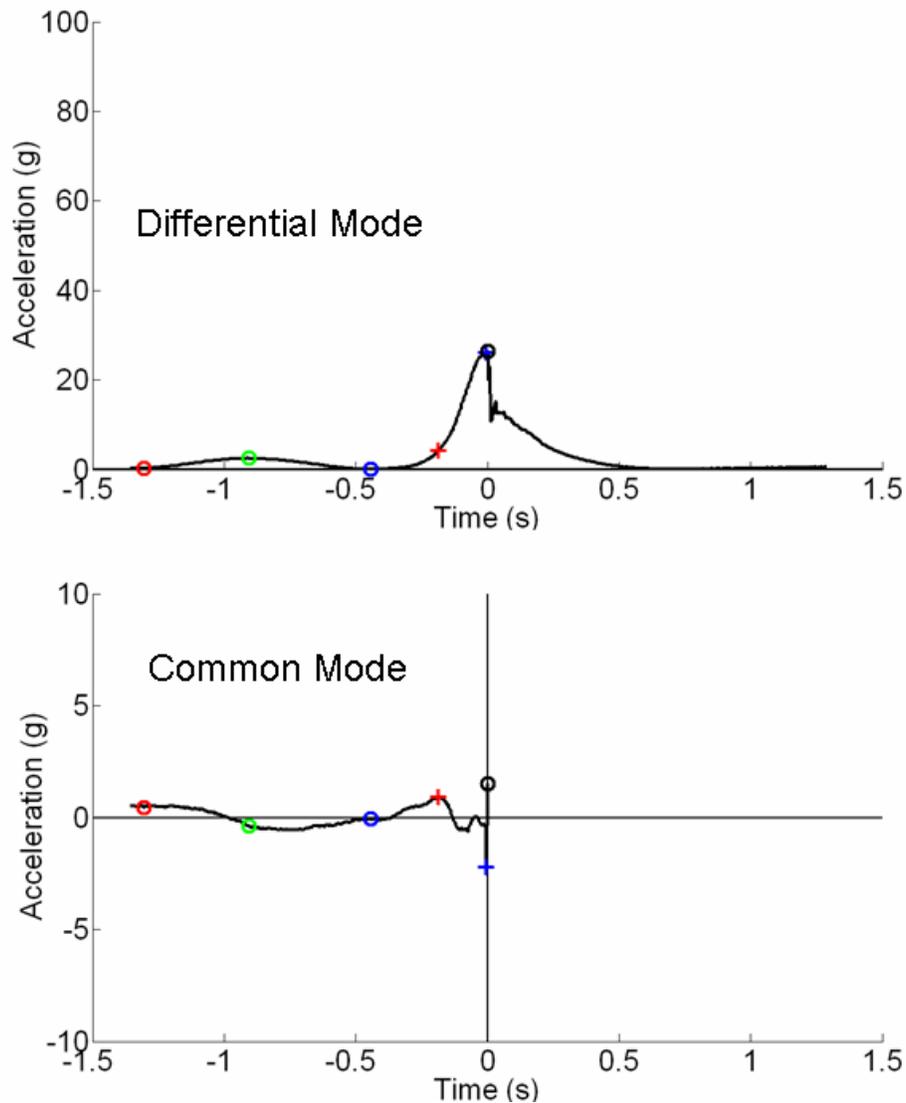

Fig. 3: Differential and common mode data averaged over seven swings of an elderly, avid, amateur golfer with a handicap of order twenty-five. The duration of the backswing is 830 ± 18 ms and the duration of the downswing is 433 ± 18 ms, which is remarkably consistent and characteristic of this beautifully rhythmic golf swing. However, this golfer is not particularly strong, as is clear from the peak of the differential mode $G(t)$ and the very small structure in the common mode $F(t)$. It is interesting that while the duration of the backswing is only slightly slower than that of the professional golfer (Fig. 6), the duration of the downswing is much slower, suggesting significantly less physical strength.

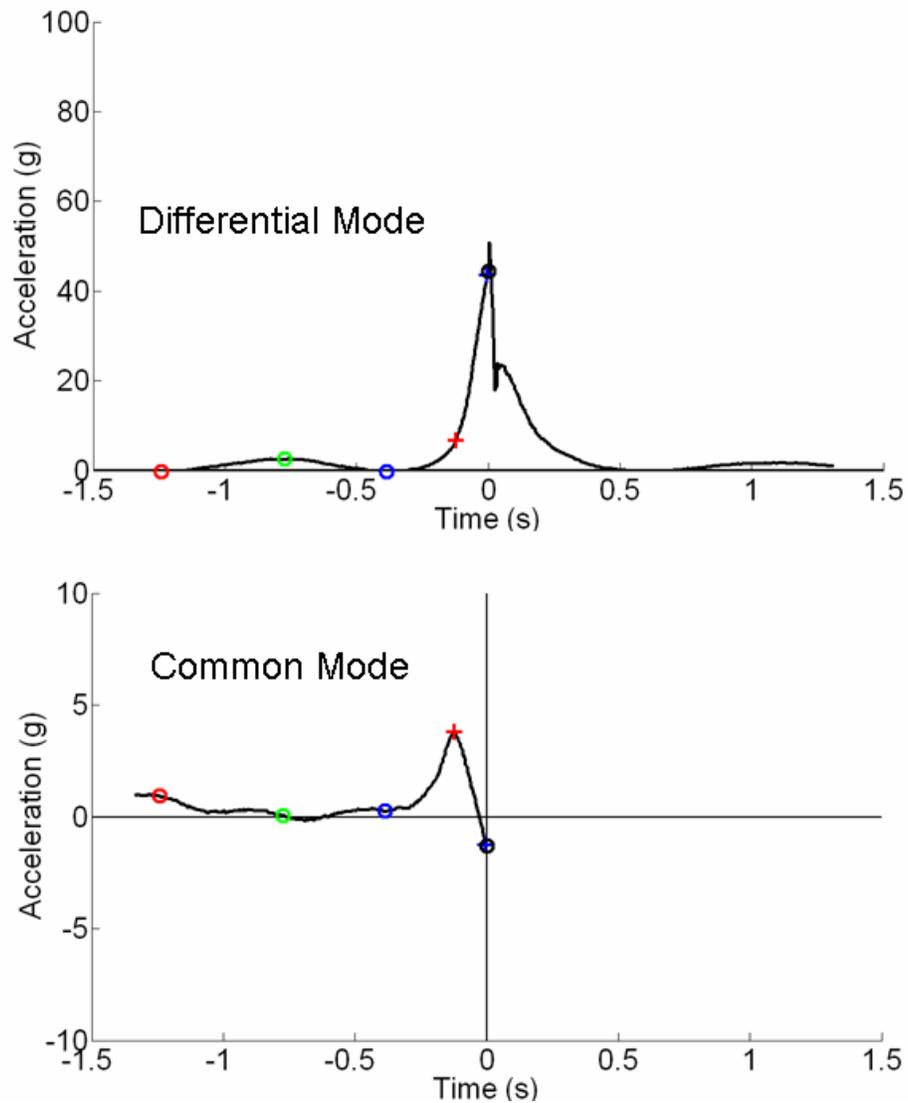

Fig. 4: Differential and common mode data averaged over six swings of an amateur golfer with a handicap of order ten. The duration of the backswing is 817 ± 16 ms and the duration of the downswing is 385 ± 20 ms. It is particularly interesting to note that while this golfer has a very respectable acceleration of the hands at the beginning of the downswing, he does not fully release the club. This is generally done so as to obtain better control over the club head, but at the cost of significant club head speed.

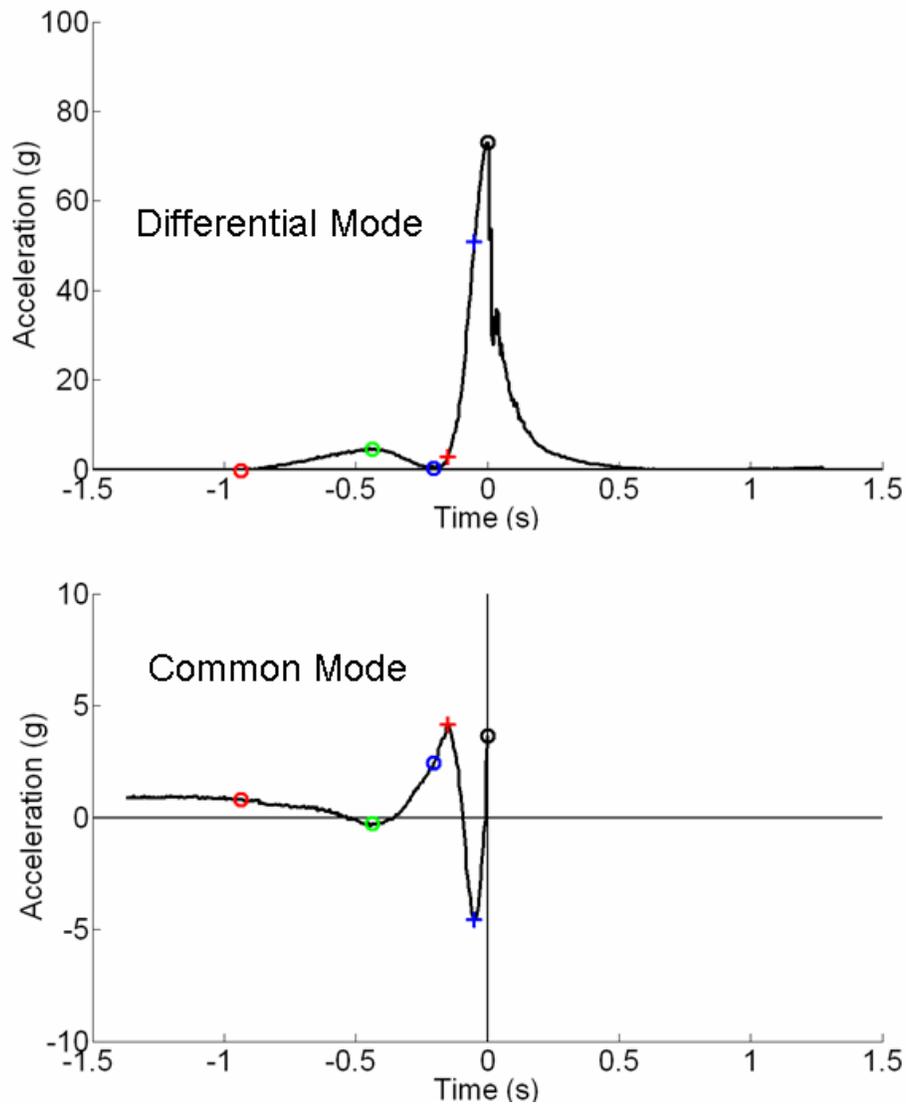

Fig. 5: Differential and common mode data averaged over four swings of a collegiate golfer. The duration of the backswing is 692 ± 48 ms and the duration of the downswing is 201 ± 33 ms, which is extremely fast but not particularly beneficial. Fluctuations in tempo are of order 10%. The club speed at impact is less than that of the professional golfer shown in the following figure. However, the common mode $F(t)$ data shows that there is clearly the beginning of a very nice common mode structure.

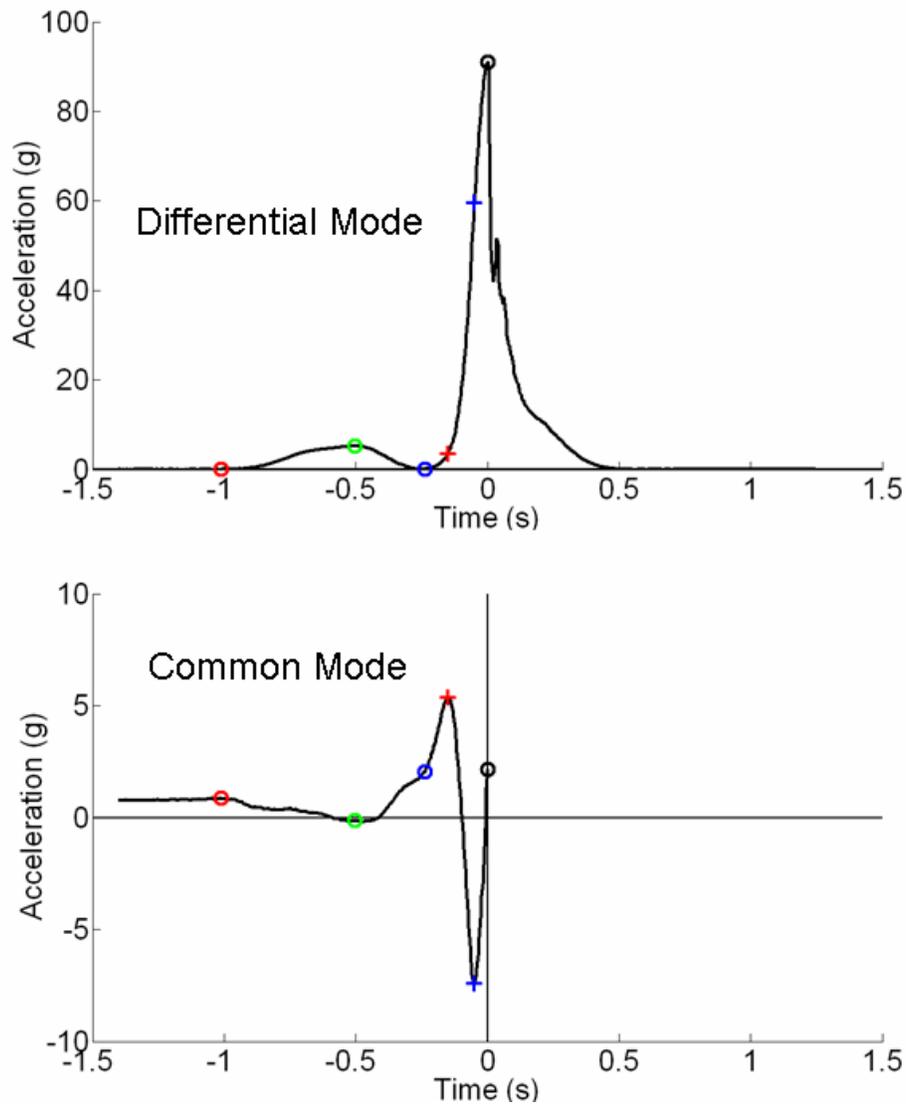

Fig. 6: Differential and common mode data averaged over eight swings of a PGA tour professional. The duration of the backswing is 724 ± 23 ms and the duration of the downswing is 248 ± 4 ms. Note that in comparison with the collegiate golfer the max-min structure in the common mode $F(t)$ is larger and wider. This implies that larger torques are sustained for longer periods of time over a larger swing arc.

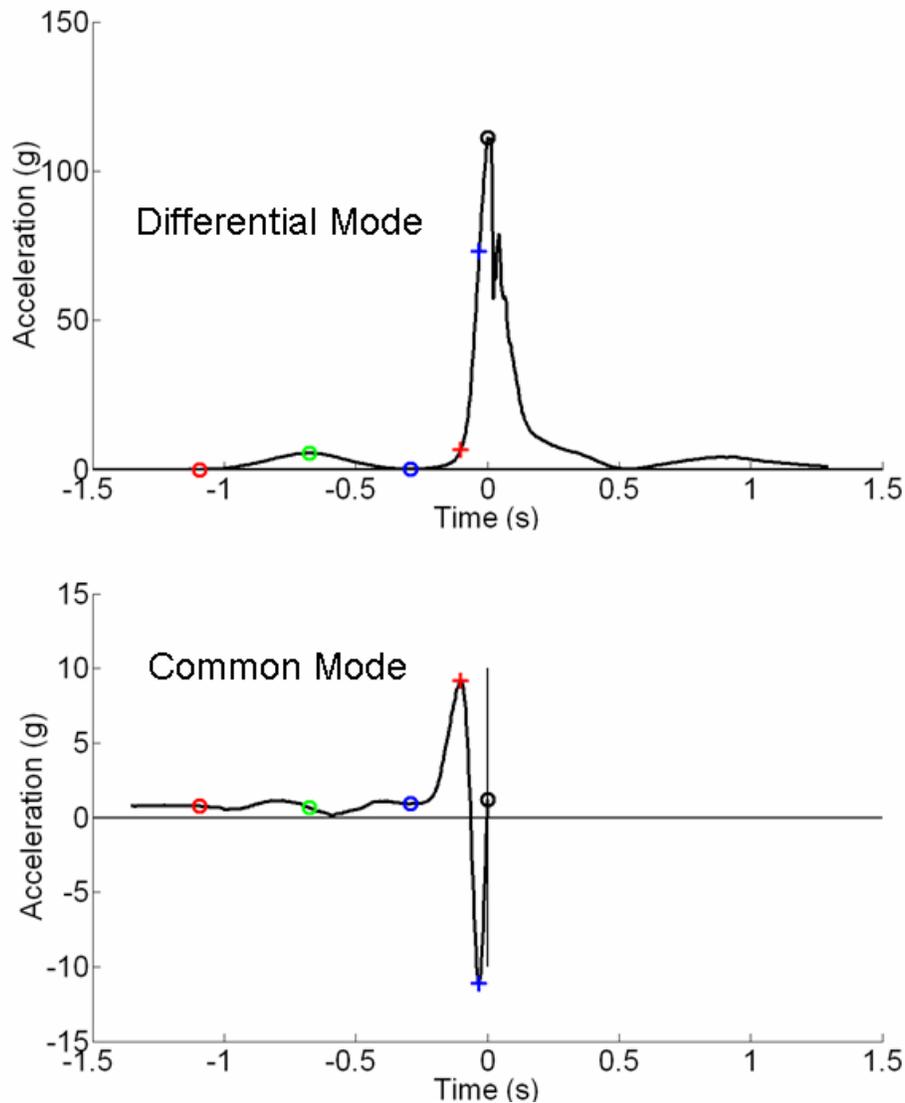

Fig. 7: Differential and common mode data averaged over eight swings of a professional long drive competitor. The duration of the backswing is 750 ± 30 ms and the duration of the downswing is 310 ± 11 ms, though it is clear that the time over which the large torques are applied in the downswing is considerably shorter. Note that **the vertical axes have been increased by 50%** relative to the other golfers in order that that the entire data set can be viewed. This golfer hits the ball far because he is very strong, generating an incredible amount of torque at the beginning of the downswing, with a peak in the common mode $F(t)$ that is of order twice the value for the tour professional.

Comparative Study of 25 Golfers

The implication of the above analysis is that the greater the max-min structure of $F(t)$ during the downswing, the greater the resulting club head speed. We have tested this hypothesis by comparing the swings of 25 golfers of varying ability, from high-handicapper to tour professional.

The test requires that we compare maximum club head speed with the size of the max-min structure of $F(t)$. The following paragraphs outline how this is accomplished.

An expression for club head speed can be derived from the expression for the position of the club head,

$$\vec{r}_c = (l_0 \cos\theta + l_c \cos\phi)\hat{x} + (l_0 \sin\theta + l_c \sin\phi)\hat{y}, \qquad (12)$$

by first taking a derivative with respect to time,

$$\dot{\vec{r}}_c = (\dot\theta l_0 \sin\theta + \dot\phi l_c \sin\phi)\hat{x} + (\dot\theta l_0 \cos\theta + \dot\phi l_c \cos\phi)\hat{y}, \qquad (13)$$

and then taking the absolute value of this velocity,

$$\left|\dot{\vec{r}}_c\right|^2 = \dot\phi^2 l_c^2 + \dot\theta^2 l_0^2 + 2\dot\theta\dot\phi l_0 l_c \cos\beta. \qquad (14)$$

In the vicinity of impact, $\beta \sim 0$, $\cos\beta \sim 1$, and the above expression simplifies to $\left|\dot{\vec{r}}_c\right| = \dot\phi l_c + \dot\theta l_0$. Now the length of the lower arm of the pendulum (i.e. the club) is longer than the length of the upper arm (i.e. the golfer's arms), $l_c > l_0$, and in the vicinity of impact $\dot\phi > \dot\theta$. With these limits, $\left|\dot{\vec{r}}_c\right| = \dot\phi l_c \left(1 + \dfrac{\dot\theta}{\dot\phi}\dfrac{l_0}{l_c}\right) \approx \dot\phi l_c$. Thus, it is not completely unreasonable to use $G(t) = (r_1 - r_2)\dot\phi^2$ as a proxy indicator for relative club head speed near to impact. In particular, we will use the peak value of $G(t)$ in the vicinity of impact.

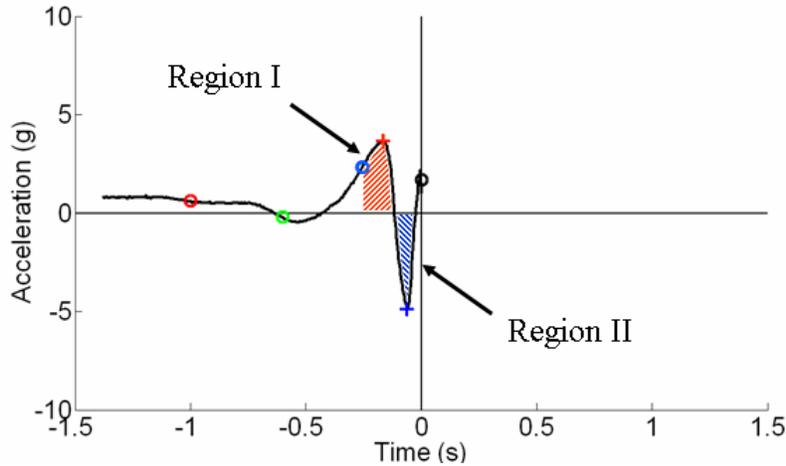

Fig. 8: The signal $F(t)$ for a golf swing. As a measure of the size of $F(t)$ during the downswing, we integrate $F(t)$ over two regions. The first region (red shaded region) starts at the beginning of the downswing (i.e blue circle), goes through the maximum, and terminates at the zero crossing. This integration yields a speed which is interpreted as approximating the maximum hand speed during the downswing. The second region (blue shaded region) spans the negative going region of $F(t)$. This integration yields a speed which is interpreted as approximating the amount by which the hands slow down as impact is approached.

For most of the downswing $\beta \sim \pi/2$, and therefore $F(t) \sim l_0 \ddot{\theta}$ is the acceleration of the hands. Integrating acceleration through time results in a change in velocity. Our proxy for the $F(t)$ max-min structure in the downswing will involve an integration of $F(t)$ through two regions. As is indicated in Fig. 8, Region I is the red shaded region. It begins at the start of the downswing, a point at which the entire system is moving very slowly. Region I continues through the peak in $F(t)$ and terminates at the zero crossing of $F(t)$. This integral yields a speed, $v_I$, which is approximately the maximum speed of the hands during the downswing. Region II is indicated as the blue shaded region and spans the entire negative region of $F(t)$. This integration yields a negative number, $v_{II}$, which is interpreted as approximating the amount by which the hands slow down as the club is released. Finally, as a measure of the size of the max-min structure associated with $F(t)$ on the downswing, we add $v_I$ to the negative of $v_{II}$, yielding $v_F = v_I - v_{II}$. While one can conceive of a myriad of other indicators, $v_F$ is particularly simple.

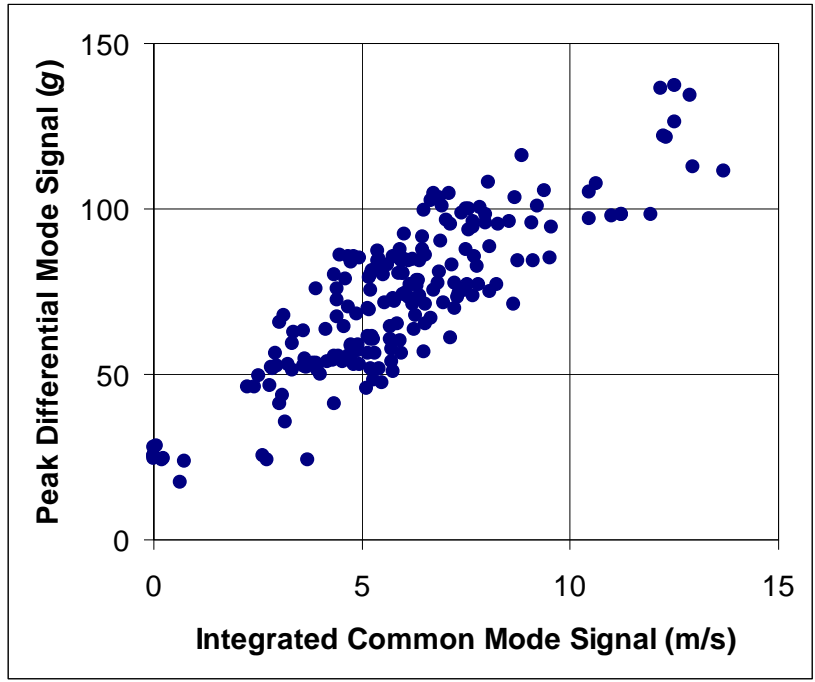

Fig. 9: The peak value of $G(t)$ as a function of $v_F$ for 200 golf swings sampled from 25 golfers. As defined in the text, $v_F$ is a proxy for the size of the max-min structure in $F(t)$ and the peak of $G(t)$ is a proxy for the club head speed at impact. The golfers range widely in capability from high-handicappers to PGA Tour professionals and professional long ball competitors. As described in the text, the correlation coefficient for this data is 0.85, indicating a high degree of correlation. Thus, we contend that the size of the structure in $F(t)$ correlates well with the speed of the club at impact.

In Fig. 9, the maximum value of $G(t)$ is plotted relative $v_F$ for 200 golf swings sampled from 25 golfers. The golfers range widely in capability from high-handicappers to PGA Tour professionals and professional long ball competitors. For each golfer, 5-10 swings are recorded while they are hitting either a 5 or 6 iron. The data indicate a clear trend: the greater is the club speed proxy, the larger is the max-min structure associated with $F(t)$. For reference, the correlation coefficient is 0.85, calculated as

$$cor = \frac{\sum (x-\bar{x})(y-\bar{y})}{\sqrt{\sum (x-\bar{x})^2 \sum (y-\bar{y})^2}} \qquad (15)$$

where $x$ is $v_F$, and $\bar{x}$ is the mean of $v_F$, $y$ is the peak of $G(t)$ and $\bar{y}$ is the mean of the peak of $G(t)$.

It is important to emphasize that while these measurements are precise, their interpretation is approximate. In particular the interpretation of the integrals $v_I$ and $v_{II}$ as indicating a change in hand speed are only accurate in the limits that the upper and lower arms of the pendulum move in the same plane, the length of the upper arm, $l_0$ does not change, $\beta \sim \pi/2$, etc. Indeed, it is likely that none of these constraints are exact, varying for every golfer and every golf swing, which perhaps accounts for the width of the distribution in Fig. 9. For instance, if a golfer lays the club off at the top of the swing, making it flat relative to the swing plane, the sensors in the shaft are not perfectly aligned with the direction of motion of the hands. In this condition, the maximum in $F(t)$ will be suppressed.

So, while there is no delusion here regarding perfect interpretation of measurement, the trend in Fig. 9 is clear. The maximum of $G(t)$, which is a very reasonable proxy for club head speed, scales with the size of the max-min structure of $F(t)$, which is a very reasonable proxy for the torques which accelerate the hands.

In summary, the common mode signal $F(t)$ provides deep insight into how torque is applied to generate club head speed in the golf swing. It is a two step process. Starting at the beginning of the downswing, the first phase involves a rapid acceleration of the hands and club. The height and width of the maximum of $F(t)$ is a measure of this initial acceleration. This is then followed by a second phase, the release, in which the club accelerates while the hands decelerate. The depth and width of the minimum of $F(t)$ is a measure of the intensity of the release.

Conclusion

The motion of the golf club has been measured using two accelerometers mounted at different points along the shaft of the golf club, both sensitive to acceleration along the axis of the shaft. Interpreted within the context of the double pendulum model of the golf swing, the resulting signals are resolved into differential and common mode components. The differential mode, a measure of the centripetal acceleration of the golf club, is a reasonable proxy for club speed and can be used to understand details of tempo, rhythm, and timing. The common mode, related to the acceleration of the hands, allows insight into the torques that generate speed in the golf swing.

A comparative study of twenty-five golfers reveals that club head speed in accomplished golfers is generated as a two step process. Starting at the beginning of the downswing, the first phase involves a rapid acceleration of the hands and club. This is then followed by a second phase, the release, in which the club accelerates while the hands decelerate. This phenomenon is clearly revealed in the common mode/differential mode data analysis.

This paper demonstrates that this measurement scheme yields a robust data set which provides deep insight into tempo, rhythm, timing, and the torques that generate power in the golf swing.

Footnotes and References:

1) See for instance, United States Patents 6,648,769; 6,402,634; 6,224,493; 6,638,175; 6,658,371; 6,611,792; 6,490,542; 6,385,559; and 6,192,323.

2) R.D. Grober, *An Accelerometer Based Instrumentation of the Golf Club: Measurement and Signal Analysis*, arXiv:1001.0956v1 [phys.ins-det].

3) T.P. Jorgensen, *The Physics of Golf* (AIP Press, American Institute of Physics, New York, 1994).

4) See United States Patent 7160200 and Sonic Golf, Inc., www.sonicgolf.com.

5) J. Novosel and J. Garrity, *Tour Tempo*, (Doubleday, New York, 2004).

6) R.D. Grober, J. Cholewicki, *Towards a Biomechanical Understanding of Tempo in the Golf Swing*, arXiv:physics/0611291v1.

7) L.D. Landau, E.M. Lifshitz, *Mechanics*, 3rd Edition, (Elsevier, New York, 1976), pg. 11.

8) W.M. Pickering and G.T. Vickers, *On the double pendulum model of the golf swing*, Sports Engineering **2**, 161-172 (1999).